\documentclass[reprint,amsmath,amssymb,aps]{revtex4-2}
\usepackage{float}
\makeatletter

\let\newfloat\newfloat@ltx
\makeatother

\usepackage{xr-hyper}
\usepackage{graphicx}
\usepackage{dcolumn}
\usepackage{bm}
\usepackage{algpseudocode}
\usepackage{siunitx}
\usepackage[english]{babel}
\usepackage{epstopdf}
\usepackage{amsmath,float}
\usepackage{amsfonts}
\usepackage{makecell}
\usepackage{booktabs}
\usepackage{threeparttable}
\usepackage{enumitem}
\usepackage{multirow}
\usepackage{hyperref}
\usepackage{natbib}
\usepackage{xcolor}
\usepackage{soul}
\newcommand{\Cov}{\mathrm{Cov}}
\begin{document}

\title{Critical nonlinear aspects of hopping transport for reconfigurable logic\\in disordered dopant networks}
\author{Henri Tertilt$^{1*}$, Jonas Mensing$^{1*}$, Marlon Becker$^{1}$, Wilfred G. van der Wiel$^{2,3}$, Peter A. Bobbert$^{3,4,5}$, Andreas Heuer$^1$}

\affiliation{
$^1$ Institute for Physical Chemistry, \\
$^2$ Department of Physics, University of M\"{u}nster, Germany \\
$^3$NanoElectronics Group, Faculty of Electrical Engineering, Mathematics and Computer Science, MESA+ Institute for Nanotechnology, and Center for Brain-Inspired Nano Systems (BRAINS), University of Twente, Enschede, The Netherlands\\
$^4$Department of Applied Physics, \\
$^5$Center for Computational Energy Research (CCER), Eindhoven University of Technology, The Netherlands\\
$^*$These authors contributed equally to the work}

\date{\today}

\begin{abstract}

Nonlinear behavior in the hopping transport of interacting charges enables reconfigurable logic in disordered dopant network devices, where voltages applied at control electrodes tune the relation between voltages applied at input electrodes and the current measured at an output electrode. 
From kinetic Monte Carlo simulations we analyze the critical nonlinear aspects of variable-range hopping transport for realizing Boolean logic gates in these devices on three levels. First, we quantify the occurrence of individual gates for random choices of control voltages. We find that linearly inseparable gates such as the XOR gate are less likely to occur than linearly separable gates such as the AND gate, despite the fact that the number of different regions in the multidimensional control voltage space for which AND or XOR gates occur is comparable. Second, we use principal component analysis to characterize the distribution of the output current vectors for the (00,10,01,11) logic input combinations in terms of eigenvectors and eigenvalues of the  output covariance matrix. This allows a simple and direct comparison of the behavior of different simulated devices and a comparison to experimental devices. Third, we quantify the nonlinearity in the distribution of the output current vectors necessary for realizing Boolean functionality by introducing three nonlinearity indicators.
The analysis provides a physical interpretation of the effects of changing the hopping distance and temperature and is used in a comparison with data generated by a deep neural network trained on a physical device.

\end{abstract}

\maketitle

\section{\label{sec:introduction}Introduction}
The development of reconfigurable logic \cite{ LeCun2015} enables new approaches for computing, using the concept of intelligent matter \cite{Kaspar2021,Menzel2021}. A key ingredient is nonlinear behavior \cite{ Maass2002,Dale2016, Tanaka2019}. A large range of physical properties can be generate by appropriate doping of semiconductors \cite{Fanciulli2021}. Here we study a dopant network processing unit (DNPU), where dopants are implanted in silicon with a concentration favoring variable-range hopping of electrons among the dopants close to the silicon surface \cite{Chen2020}. The dopant network is contacted by electrodes deposited on the silicon surface, allowing the application of voltages and the measurement of currents. Reconfigurable logic functionality can be obtained with these DNPUs, among which Boolean functionality.  The standard usage is at liquid-nitrogen temperature ($T=77$ K), but room-temperature operation was also demonstrated \cite{Chen2020}. Increasingly complex functionalities can be achieved by interconnecting DNPUs in networks \cite{RuizEuler2021}. Logic circuitry based on DNPUs has the potential to outcompete CMOS-based logic in terms of energy efficiency, latency and footprint \cite{Chen2020}.

We consider here a DNPU with eight contacts: one electrode is chosen as output, two electrodes are chosen as inputs, while voltages applied at the other five electrodes control the input-output characteristics. To verify whether the system behaves like one of the six basic Boolean logic gates (AND, OR, NOR, NAND, XOR, XNOR), one applies voltages to the two input electrodes corresponding to the possible logic combinations `00', `10', `01', `11'. By studying the four-dimensional vector of currents at the output electrode for these inputs one can check whether the system displays the desired Boolean functionality. By introducing a fitness function for each logic gate, the functionality can be quantified and subsequently optimized by variation of the voltages at the control electrodes. Two different approaches for this optimization have been used: a genetic algorithm \cite{Bose2015,Chen2020} and gradient descent on a deep-neural-network (DNN) surrogate model (SM) trained to reproduce the full input-output characteristics of the DNPU \cite{RuizEuler2020}.


In order to obtain a detailed physical understanding of the atomic-scale behavior of the DNPUs, we recently developed a microscopic model of their functioning  \cite{Tertilt2022}. The model is based on variable-range hopping of interacting charges. The charge hopping was simulated with kinetic Monte Carlo (KMC) simulations. Like in the experiment, realizations of Boolean logic gates were obtained by varying the control voltages. We gained important insight into the functioning of DNPUs by mapping out the spatial current and voltage distribution for high-fitness realizations of specific gates. By studying the sensitivity of the fitness to variations of the control voltages, the impact of nonlinear effects and the major differences between the linearly separable AND gate and the linearly inseparable XOR gate were analyzed.

In the present work, we take a complementary approach to understanding the functioning of DNPUs by analyzing their statistical properties and identifying the critical nonlinear aspects that allow for reconfigurable logic.
We study the four-dimensional (4D) current vectors for the input combinations `00', `10', `01', `11' for a large number of different control voltage combinations. Experimentally, a similar study was done, giving rise to abundance plots of individual Boolean gates as a function of a minimal gate fitness \cite{Chen2020}. Here, we extend and go beyond the analysis of the abundance plots with the goal to grasp the key statistical features of the distribution of Boolean gate realizations in the five-dimensional (5D) space of control voltages. 

As a central approach, we use a principal component analysis (PCA) \cite{Jolliffe2018} to define an orthogonal system of directions in the 4D current vector space with uncorrelated properties, resulting from an eigenvector analysis of the covariance matrix of the output current vectors. The corresponding eigenvalues contain information about the fluctuations of the current vectors in the eigenvector directions. A common application of the PCA is the simplification of high-dimensional trajectories, where directions with small eigenvalues are eliminated \cite{Antoniou2011}. Also in the field of reservoir computing the PCA method has been used, either to construct an autoencoder \cite{Penkovsky2018} or to use it as a tool to assess the internal representation ability of the self-organized reservoir. Also in that application only the directions with the largest eigenvalues matter.
By contrast, in the PCA applied in this work all eigenvector directions are important. We will see that even the direction with the smallest eigenvalue turns out to be essential for the logic functionality of DNPUs. We will also show that the PCA eigenvectors and eigenvalues allow an objective comparison with experimental data. 

Going beyond the PCA, we map the current vectors onto three variables that quantify the different nonlinear effects inherently present in the DNPU. This decomposition separates nonlinear single-input responses and nonlinear cross-input responses. Related to these three variables we introduce three key indicators that characterize the occurrence of nonlinearities in the current vector distributions. From the insights of the PCA analysis and this decomposition, key properties of the abundance plots can be understood on a deeper level than before. Specifically, we use these insights to better understand the dependence of the DNPU logic functionality on the hopping distance and on temperature. We expect that the introduced concepts have a general applicability to a large variety of nonlinear disordered systems other than DNPUs, including nanoparticle networks \cite{Bose2015}.

Additionally, we analyze the spatial correlation of fitness values for different gates in the 5D space of control voltages. In this way we obtain information about the hypervolume of individual gates in this space. This hypervolume  is directly related to the sensitivity of the gate fitness to variations of the control voltages, which is of major practical relevance.

The outline of this work is as follows. First, we introduce the theoretical background of the different concepts and models used. Then we discuss the results at the three different analysis levels mentioned above. We start with results from a general statistical analysis, in particular abundance plots and the spatial correlation of fitness values. After that we discuss the results for the output covariance matrices in the framework of the PCA. We then discuss the results obtained from the decomposition of the current vectors into the three nonlinearity indicators. Finally, we provide a summary, conclusions and an outlook.

\section{\label{sectheory}Theoretical background\protect}

\subsection{\label{sec:model}Model\protect}

\begin{figure}[b]
    \centering
    \includegraphics[width=0.45\textwidth]{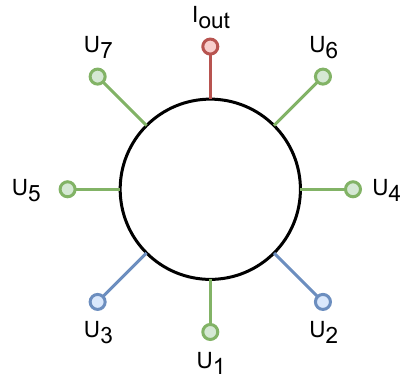}
    \caption{Sketch of the studied DNPU with 8 electrodes. Voltages $U_1$ and $U_4$-$U_7$ applied to five control electrodes tune the relation between the output current $I_{\rm out}$ and the input voltages $U_2$  and $U_3$, such that a desired logic functionality is obtained. }
    \label{fig:device}
\end{figure}

\noindent We study DNPUs with an electrode configuration as sketched in Figure~\ref{fig:device}. The input voltages are $U_3$ (also denoted as $U_{\rm in,l}$, l: left) and $U_2$ (also denoted as $U_{\rm in,r}$, r: right). The control voltages are $U_1$, and $U_4$-$U_7$. The output electrode is grounded ($U_{\rm out}=0$) and the output current is $I_{\rm out}$.  We study in this work two devices D1 and D2 with different configurations of 200 randomly placed boron dopants and 3 counterdopants in a circular area with a diameter of 150 nm. These values are representative for the experimental situation \cite{Chen2020,Tertilt2022}.

We consider electron hops between dopants $i$ and $j$, separated by a distance $r_{ij}$. We assume phonon-mediated variable-range hopping described by the Miller-Abrahams rate \mbox{$\Gamma_{ij}=\nu_0 \exp\left(-2r_{ij}/a-\Delta E_{ij}/k_{\rm B}T\right)$} if \mbox{$\Delta E_{ij}>0$} and \mbox{$\Gamma_{ij}=\nu_0 \exp\left(-2r_{ij}/a\right)$} otherwise \cite{Miller1960}. For the hopping prefactor we take a typical phonon frequency \mbox{$\nu_0=10^{12} \text{ s}^{-1}$}, but the specific value does not matter for the statistical results reported in this work. Furthermore, $k_{\rm B}T$ is the thermal energy, $\Delta E_{ij}$ the energy change involved in the hop, and $a$ is the dopant wave function decay length, which we will call the `hopping distance'. The Miller-Abrahams rate has initially been derived for electron hopping between dopants in semiconductors and it is assumed here to be valid for all hopping processes, also for for hopping between the electrodes and the boron dopants. The electrodes are regarded as infinite reservoirs of electrons and are modeled as circular segments, with their centers defining the distance to the dopants.

The following energy contributions are present: (i) the electrostatic energy, given by the external electrostatic potential imposed by the voltages applied to the electrodes, (ii) the Coulomb interactions in between all the electrons and between the electrons and the ionized counterdopants, and (iii) a random Gaussian contribution to the dopant ionization energy with a standard deviation $\sigma = 0.1$ eV, which was found in Ref.~\cite{Tertilt2022} to yield a resistance-temperature dependence for a low voltage applied between neighboring electrodes in fair agreement with experiment. To obtain the electrostatic energy we solve the two-dimensional (2D) Laplace equation for the electrostatic potential. For this, we use a triangular mesh and a finite element method based on the MFEM library \cite{MFEM}. 

The dielectric constant in the Coulomb interaction is chosen as $\epsilon_{\rm r}=12$, close to that of silicon. The hopping distance $a$ is varied from 2.5 to 10 nm for simulations at the standard temperature 77 K and from 1.25 to 10 nm for simulations at room temperature (293 K). For the comparison of the model results with experiment, we take the typical value $a=5$ nm applicable to a donor like boron in silicon.

\subsection{\label{sec:KMC}Kinetic Monte Carlo simulations\protect}

We use a standard, in-house developed rejection-free KMC algorithm that considers at each step all possible electron hops in the system \cite{Tertilt2022}. Voltages are applied to seven electrodes and the current is determined at a chosen grounded output electrode by counting the net number of electron hops to or from that electrode in the simulated time interval. Starting with as many electrons in the system as counterdopants (neutral system), $10^4$ KMC equilibration steps are sufficient to reach a steady state current for all considered voltage combinations. Unless stated otherwise, we determine the current in a time interval corresponding to \mbox{$10^7 \text{ KMC steps}$} and estimate the statistical uncertainty from the current fluctuations in 100 equally long subintervals of $10^5$ KMC steps.

\subsection{\label{sec:hypercube}Hypercube sampling\protect}

We randomly draw  control voltages from a hypercube ({\it hypercube sampling}) such that each control voltage ranges between $-1$ and 1 V. The input voltages are either \mbox{0 V (logical $0$)} or \mbox{0.5 V (logical $1$)}. For a given set of control voltages we obtain the four-dimensional current vector \mbox{$(I_{00}, I_{10}, I_{01}, I_{11})$} from the KMC simulations. The hypercube sampling involved in total $10^4$ different random choices of control voltages. 

\subsection{\label{sec:DNN}Comparison with deep-neural-network surrogate model\protect}

We compare results of KMC simulations with those of a deep-neural-network (DNN) surrogate model (SM) trained on experimental data \cite{RuizEuler2020}. The SM accurately reproduces the measured output current $I_{\rm out}$ as a function of all seven input voltages $U_1$-$U_7$ for an experimental DNPU as sketched in Figure~\ref{fig:device}. Hence, our comparison is equivalent to that with a real-world device. The DNN consists of an output layer with a single neuron giving $I_{\rm out}$ as output and an input layer with seven neurons for $U_1$-$U_7$ as inputs. In between are six hidden layers, each with $90$ neurons. The DNN was trained on the experimental data for voltages $U_1$-$U_5$ in the interval $[-1.2,0.6]$ V and $U_6,U_7$ in the interval $[-0.7,0.3]$ V. The comparison with the KMC simulations was done for $U_1$-$U_5$ in the interval $[-0.5,0.5]$ V and $U_6,U_7$ in the interval $[-0.3,0.3]$ V to avoid extrapolation beyond the trained range. The input voltages for Boolean logic in the comparison are either 0 V (logical 0) or 0.1 V (logical 1).

\subsection{\label{sec:DNN}Fitness function\protect}

For the abundance plots of a given Boolean logic gate we have used the fitness function $F$ defined as \cite{Tertilt2022,Chen2020}
\begin{equation}\label{eq:fitnessfunction}
F = \frac{m}{\sqrt{\rm MSE} + k|c|},
\end{equation}
where $m$ and $c$ are fit parameters of a linear fit
\mbox{$I_{\rm out} (U_{\rm in,l}, U_{\rm in,r}) = m \, G(U_{\rm in,l}, U_{\rm in,r}) + c$}. Here, \mbox{$G(U_{\rm in,l}, U_{\rm in,r})$} is the logic table of the considered gate. MSE denotes the mean squared error of the linear regression. For the constant $k$ we choose 0.01, as in the experimental work \cite{Chen2020}. A finite value of $k$ rewards a large relative separation of the high and low current levels, which is relevant for the experimental separation of these levels. However, as discussed below, when applying the PCA, normalized currents have to be used, which are not sensitive to this separation.

\subsection{\label{sec:PCA}Data preparation \protect}

From the hypercube sampling we obtain a set of $10^4$ current vectors \mbox{$(I_{00}, I_{10}, I_{01}, I_{11})$}. The fitness values used in the abundance plots are directly based on this data set. For the subsequent analysis we transform each current vector by subtracting the average current $I_{\rm av}=\frac{1}{4}\left(I_{00}+I_{10}+I_{01}+I_{11}\right)$ of the four components from each component. First, in this way we increase the sensitivity to nonlinear effects relative to the average current and, second, the analytical calculations, outlined below, can be performed by solving quadratic rather than cubic equations. We denote the average of $I_{\rm av}$ over all $10^4$ current vectors as $\langle I_{\rm av}\rangle $.

\subsection{\label{sec:PCA}Principal component analysis \protect}

In a principal component analysis (PCA), fluctuations in a multi-component variable are expressed along orthogonal directions that are linearly uncorrelated. The first direction displays the largest fluctuations. After projecting out the first direction, the second direction displays the largest fluctuations in the remaining subspace, and so on. In many applications the PCA is used to reduce the dimension of the problem at hand. For example, projecting out the dimension with the lowest eigenvalue of the covariance matrix typically has limited impact on the data set, but allows for a lower-dimensional description. Here, we use the PCA to characterize the statistical fluctuations in the set of current vectors \mbox{$(I_{00}, I_{10}, I_{01}, I_{11})$} obtained in the hypercube sampling. This allows for the identification of relevant directions and inherent symmetries as well as 
a direct comparison between simulation and experiment.

To apply the PCA, we consider the set of $10^4$ vectors \mbox{$(I_{00}, I_{10}, I_{01}, I_{11})$} from the hypercube sampling (after subtracting $I_{\rm av}$ from each component). From this set one can define the symmetric covariance matrix $C$:
\begin{widetext}
    \begin{equation}
        \label{eq:covariancematrix}
        C=
        \begin{pmatrix}
        \sigma^2 (I_{00}) & \Cov(I_{00}, I_{10}) & \Cov(I_{00}, I_{01}) &\Cov(I_{00}, I_{11})\\ \Cov(I_{10}, I_{00})  & \sigma^2 (I_{10}) & \Cov(I_{10}, I_{01}) & \Cov(I_{10}, I_{11}) \\ \Cov(I_{01}, I_{00})  & \Cov(I_{01}, I_{10}) & \sigma^2 (I_{01}) & \Cov(I_{01}, I_{11}) \\ \Cov(I_{11}, I_{00})  & \Cov(I_{11}, I_{10})  & \Cov(I_{11}, I_{01})  &  \sigma^2 (I_{11}) 
        \end{pmatrix},
    \end{equation}
\end{widetext}
where $\sigma^2\left(I_{ij}\right)$ are the variances and ${\rm Cov}\left(I_{ij},I_{kl}\right)$ the covariances of the current vector components.
The PCA implies diagonalization of $C$, yielding the eigenvalues $\lambda_i$ and the corresponding eigenvectors $\mathbf{J}_i$ ($i=0,...,3$). Due to the subtraction of $I_{\rm av}$ from each component, one eigenvalue is $\lambda_0=0$ with eigenvector $\mathbf{J}_0=\frac{1}{2} (1,1,1,1)$. The remaining eigenvalues are sorted such that $\lambda_j \le \lambda_i $ if $j > i$.

\subsection{\label{sec:Dec}Decomposition procedure \protect}

In a decomposition procedure that turns out to be very useful, we map the current vectors \mbox{$(I_{00}, I_{10}, I_{01}, I_{11})$} onto four new variables via
\begin{equation}
\label{eq:vardef}
\begin{array}{r@{}l}
I_{\rm av}  &{} = \frac{1}{4} (I_{11} + I_{10} + I_{01} + I_{00} ), \\
M_{\rm l}   &{} = \frac{1}{4} (I_{11} + I_{10} - I_{01} - I_{00} ), \\
M_{\rm r}   &{} = \frac{1}{4} (I_{11} - I_{10} + I_{01} - I_{00} ), \\
X           &{} = \frac{1}{4} (I_{11} - I_{10} - I_{01} + I_{00} ).
\end{array}
\end{equation}
These variables have the following interpretation: (i) $I_{\rm av}$ is the average current introduced above (it is zero if $I_{\rm av}$ was already subtracted from the current vector components). (ii) $M_{\rm l}$ reflects the increase of $I_{\rm out}$ upon increasing the voltage of the left electrode. This increase is averaged over the two possible input voltages of the right electrode. $M_{\rm l}$ can thus be interpreted as an effective conductance with respect to the left input voltage. (iii) $M_{\rm r}$ has a similar interpretation as $M_{\rm l}$ but with respect to the right electrode. (iv) In case that the increase of $I_{\rm out}$ upon increasing the left input voltage is independent of the right input voltage one has $I_{11} - I_{01} = I_{10} - I_{00}$. This is equivalent to $X = 0$. An equivalent argument holds when left and right electrode are interchanged. Thus,  $X$ is a measure for the cross-correlation between the two inputs and can thus be interpreted as a nonlinear coupling between them.

\subsection{\label{sec:PCA-E}Calculation of PCA eigenvalues and eigenvectors \protect}

In terms of the new variables, Eq.~(\ref{eq:vardef}) can be rewritten as

\begin{equation}
\label{eq:IvsO}
\begin{array}{r@{}l}
I_{00} &{} = I_{\rm av} - M_{\rm l} - M_{\rm r} + X,\\
I_{10} &{} = I_{\rm av} + M_{\rm l} - M_{\rm r} - X,\\
I_{01} &{} = I_{\rm av} - M_{\rm l} + M_{\rm r} - X,\\
I_{11} &{} = I_{\rm av} + M_{\rm l} + M_{\rm r} + X.
\end{array}
\end{equation}
In principle, the eigenvectors and eigenvalues of the PCA covariance matrix can be expressed in terms of the variances and covariances of the new variables appearing on the right hand side of Eq.~(\ref{eq:IvsO}). To simplify the calculation we will neglect correlations between $X$ and the two variables $M_{\rm l}$ and $M_{\rm r}$. As shown below, the corresponding Pearson correlation coefficients are indeed very small.
Then, each term of the PCA matrix can, for $I_{\rm av}=0$, be written in the form
\begin{eqnarray}
    \Cov(I_{ij},I_{kl} ) &=&  d_{\rm l} \sigma^2( M_{\rm l}) + d_{\rm r}  \sigma^2( M_{\rm r})\nonumber\\
    &+& d_x  \sigma^2 (X) + 2d_{\rm lr}  \Cov(M_{\rm l}, M_{\rm r}),
\end{eqnarray}
with $\{i,j,k,l\} \in \{0,1\}$, $\{d_{\rm l},d_{\rm r},d_x\} \in \{-1,1\}$, and  $\{d_{\rm lr}\} \in \{-1,0,1\}$. The calculation of these covariances is straightforward. E.g., one has $\Cov(I_{00},I_{01}) = \sigma^2( M_{\rm l}) - \sigma^2( M_{\rm r}) - \sigma^2 (X) + 2\times 0 \times  \Cov(M_{\rm l}, M_{\rm r})$.

Next, we define the four vectors
\begin{equation}
\begin{array}{r@{}l}
\mathbf{v}_0&{}=\frac{1}{2} (1,1,1,1),\\
\mathbf{v}_1&{}=\frac{1}{\sqrt{2}} (0,-1,1,0),\\ 
\mathbf{v}_2&{}=\frac{1}{\sqrt{2}} (-1,0,0,1),\\
\mathbf{v}_3&{}=\frac{1}{2} (-1,1,1,-1).
\end{array}
\end{equation}
One can directly check that $\mathbf{J}_0=\mathbf{v}_0$ and $\mathbf{J}_3=\mathbf{v}_3$ are eigenvectors of the PCA covariance matrix with eigenvalues $\lambda_0 = 0$ and $\lambda_3 = 4 \sigma^2(X)$, respectively. The two remaining eigenvalues can be written as
\begin{eqnarray}
    \lambda_{1,2} &=& 2(\sigma^2( M_{\rm l}) + \sigma^2(M_{\rm r}) )\nonumber\\
        &\pm& \sqrt{(\sigma^2(M_{\rm l}) - \sigma^2( M_{\rm r}))^2 + 4 [\Cov( M_{\rm l},M_{\rm r})]^2}\nonumber
\end{eqnarray}
The corresponding eigenvectors $\mathbf{J}_1$ and $\mathbf{J}_2$ are linear combinations of $\mathbf{v}_1$ and $\mathbf{v}_2$. 

The result becomes particularly simple if one has $\sigma^2(M_{\rm l}) =\sigma^2(M_{\rm r}) \equiv \sigma^2 (M_{\rm l,r})$. We denote this scenario as {\it l-r symmetry}. A special realization of l-r symmetry occurs when the arrangement of dopants and electrodes displays left-right mirror symmetry (up-down mirror symmetry in Figure~\ref{fig:device}), which in practice is only an idealized limit. The eigenvectors are then given by $\mathbf{J}_i=\mathbf{v}_i$ with eigenvalues
\begin{equation}\label{eq:eigenvalues}
\begin{array}{r@{}l}
 \lambda_0 &{} =  0, \\   
  \lambda_1 &{} =  4 \sigma^2(M_{\rm l,r}) (1 + {\rm Corr}(M_{\rm l},M_{\rm r})), \\   
   \lambda_2 &{} =  4 \sigma^2(M_{\rm l,r}) (1 - {\rm Corr}(M_{\rm l},M_{\rm r})),\\   
    \lambda_3 &{} =  4 \sigma^2(X).  
\end{array}
\end{equation}
where ${\rm Corr}(A,B) $ is the Pearson correlation coefficient between $A$ and $B$. 

\subsection{\label{sec:Dec} Nonlinearity indicators \protect}

\begin{figure*}[t]
    \includegraphics[width=\textwidth]{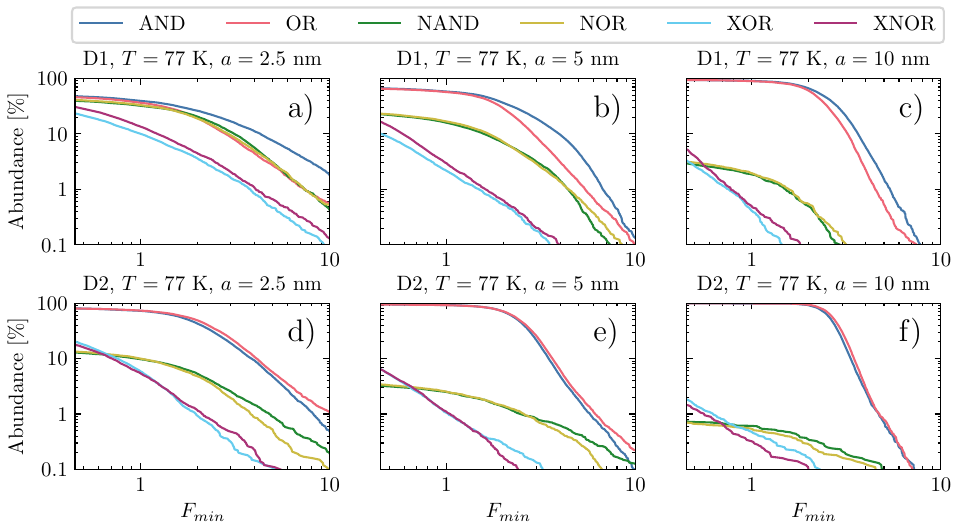}
    \caption{\label{fig:device_abundance_comparsion}Abundance at $T = 77$ K of Boolean gates with fitness higher than a fitness threshold $F_\mathrm{min}$ in the 5D space of control voltages for the two devices D1 and D2 and hopping distance $a=2.5$, $5$ and $10$ nm.}
\end{figure*}

From the hypercube sampling one obtains distributions of the three variables $M_{\rm l}$, $M_{\rm r}$ and $X$. We will consider here the first moment $\langle A\rangle$, the second moment $\langle A^2 \rangle $, and the variance $\sigma^2(A)$ of these distributions. If the dopant network would show a purely linear response to changes in the input voltages, $I_{1j} - I_{0j}$ would be a constant $2M_{\rm l}^0$ independent of $j$ and independent of the control voltages. For the probability distribution function of $M_{\rm l}$ we would then have $p(M_{\rm l}) = \delta (M_{\rm l} - M_{\rm l}^0)$. The same would hold for the probability distribution function of $M_{\rm r}$: $p(M_{\rm r}) = \delta (M_{\rm r} - M_{\rm r}^0)$. $M_{\rm r}^0$ could be different from $M_{\rm l}^0$, e.g., if the positions of the dopants would not be fully symmetric relative to the two input electrodes. Due to nonlinear effects it is expected that $M_{\rm l}$ and $M_{\rm r}$ fluctuate for the different control voltages chosen in the hypercube sampling. For perfect realizations of Boolean gates we have $I_{01} =  I_{10}$, corresponding to $M_{\rm l} = M_{\rm r}$. This is automatically fulfilled if ${\rm Corr}(M_{\rm l},M_{\rm r}) =1$. Thus, one may expect that a high Pearson correlation between $M_{\rm l}$ and $M_{\rm r}$ is advantageous for the realization of all gates.

For the realization of NAND and NOR gates, it is essential that an increase of an input voltage may lead to a decrease of the output current. This is equivalent to the occurrence of a negative differential resistance (NDR). In our notation, this would imply $M_{\rm l} < 0$ (and/or $M_{\rm r} < 0$). Thus, of utmost relevance is the probability that NDR occurs. For this purpose we define the left and right NDR indicators 
\begin{eqnarray}\label{eq:QlQr}
    Q_{\rm l} &\equiv& \frac{1}{2} \left [1 - \tanh \left (  \frac{\langle M_{\rm l}\rangle }{\sigma(M_{\rm l})} \right )  \right ] \nonumber \\
    Q_{\rm r} &\equiv& \frac{1}{2} \left [1 - \tanh \left (  \frac{\langle M_{\rm r}\rangle }{\sigma(M_{\rm r})} \right )  \right ].  
\end{eqnarray}
For a symmetric distribution, a value of $Q_{\rm l} = 0.5$ implies that in 50\% of all realizations NDR is present for the left input, because the average $\langle M_{\rm l} \rangle$ is then zero. In contrast, in the limit  $Q_{\rm l} \rightarrow 0$ the first moment is positive and much larger than the standard deviation. Thus, NDR does not occur. Analogously, in the opposite limit $Q_{\rm l} \rightarrow 1$ NDR will always occur. In general, the larger $Q_{\rm l}$, the higher the probability that NDR occurs for a certain combination of control voltages. Thus, $Q_{\rm l}$ is a measure of how likely NDR is upon variation of the left input voltage. Identical arguments hold for $Q_{\rm r}$. Note that in this extension of the PCA also the first moment of the distributions plays an essential role.

Both the XOR and the XNOR gates are linearly inseparable. Among others things, this implies that non-monotonic behavior upon increasing the sum of the two input voltages must be present. The occurrence of such non-monotonic behavior is strongly connected to the variable $X$. For a perfect XOR or XNOR gate one has $I_{11} = I_{00}$ and $I_{01} = I_{10}$, and thus $2X =I_{11} - I_{01}$ (or, alternatively, $2X =I_{11} - I_{10}$). A large negative value of $X$ is required for a high-fitness XOR gate and a large positive value is required for a high-fitness XNOR gate, relative to the scale of fluctuations of $I_{11} - I_{01}$ (or $I_{11} - I_{10}$). Since the typical scale of the fluctuations of $I_{11} - I_{01}$ (or $I_{11} - I_{10}$) is the same as that of $M_{\rm l}$ (or $M_{\rm r}$) and since distributions are characterized by their second moments, we choose as an indicator for the nonlinear coupling between the two inputs
\begin{equation}\label{eq:Qlr}
Q_{\rm lr} \equiv \frac{2\langle X^2\rangle }{\langle  M_{\rm l}^2 \rangle + \langle M_{\rm r}^2 \rangle}.
\end{equation}
This completes the set of three nonlinearity indicators.

\section{\label{sec:main}Results: general statistical properties\protect}
\subsection{\label{sec:main}Gate abundances\protect}

In the experimental work \cite{Chen2020} the emergence of Boolean functionality was illustrated by abundance plots, representing the probability that the current vector in a random hypercube sampling has a fitness larger than a given value $F_{\rm min}$.
In Figure~\ref{fig:device_abundance_comparsion} we show simulated abundance plots for all six Boolean gates at $T=77$~K for both devices D1 and D2, based on the random sampling of $10^4$ control voltage combinations. In order to study the influence of the hopping distance $a$, we show results for $a=2.5$, $5$ and $10$ nm. The abundance plots are qualitatively similar for both devices, but show important quantitative differences, caused by the different locations of the (counter)dopants. A general observation is that, by chance, there are less high-fitness gates for D2.

We clearly observe a similar fitness threshold-dependence of the AND and OR gates, the NAND and NOR gates, as well as the XOR and XNOR gates. This pair-wise similarity is in line with our above observation  that the values of $Q_\mathrm{l}$ and $Q_\mathrm{r}$ should be important for the realization of  NAND and NOR gates, while the value of $Q_{\mathrm{lr}}$ should be important for the realization of XOR and XNOR gates. We see in the middle panels of Figure~\ref{fig:device_abundance_comparsion} that for the standard hopping distance $a=5$ nm it is much more likely to find, for example, an AND gate than an XOR gate for a random choice of control voltages.  

We also see from Figure~\ref{fig:device_abundance_comparsion} that the number of logic gates with high fitness values strongly increases with decreasing $a$. For example, for a fitness threshold $F_\mathrm{min}=8$ the number of AND gates is increased by more than an order of magnitude for both devices when comparing $a=10$ to $a=2.5$ nm. Furthermore, the fitness distributions of the 6 different gates tend to approach each other with decreasing $a$. We attribute the increase in the occurrence of high-fitness gates with decreasing $a$ to the increased importance of Coulomb interactions in determining the current flow. For large $a$ charges can hop far away to energetically favorable sites that they cannot reach for small $a$. In this way they can bypass sites close to other charges that are inaccessible because of Coulomb repulsion. In the extreme case of very large $a$ the dopant network would act as a linearly resistive medium, with a trivial linear relation between the input voltages and the output current. It is the complex input-output relation for small $a$ due to Coulomb interactions that leads to the occurrence of high-fitness gates in the 5D control voltage space.  
We note in passing that a further decrease of $a$ below 2.5 nm results in a significant number of simulation runs where no output current is obtained on the typical time scale of the KMC simulations. This is a consequence of the exponential dependence of the hopping rate on $a$.

In Figure~\ref{fig:device_abundance_comparsion_k} of the Supplemental Material \cite{supp} we show the same data as in Figure~\ref{fig:device_abundance_comparsion}, but with $k=0$ in the definition of the fitness function Eq.~(\ref{eq:fitnessfunction}). We clearly observe that for small $F_{\rm min}$ the abundance plots are basically identical, whereas for high $F_{\rm min}$ at least a qualitative agreement remains. This is an important observation because the interpretation of the covariance matrix is particularly simple for current vectors that have been shifted by subtracting $I_\mathrm{av}$ from the current components. Since the shift automatically yields $c=0$ in the definition of the fitness function Eq.~(\ref{eq:fitnessfunction}), the shift is equivalent to choosing $k=0$ for the non-shifted currents. We may thus conclude that the fitness properties after shifting the currents hardly change.

\subsection{\label{sec:volume }Hypervolumes and numbers of gate realizations\protect}

\begin{figure*}[t]
    \includegraphics[width=\textwidth]{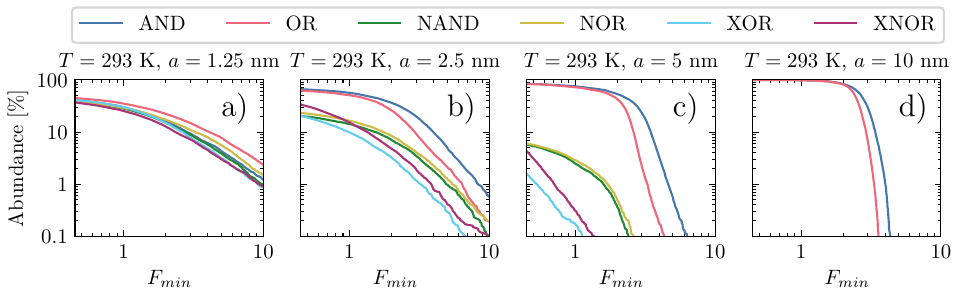}
    \caption{\label{fig:abundance_comparsion_highT}
    Abundance at $T = 293$ K of Boolean gates with fitness higher than a fitness threshold $F_\mathrm{min}$ in the 5D control voltage space for device D1 and different values of the hopping distance $a$.}
\end{figure*}

Next, we elucidate the origin of the observed major differences between AND and XOR gates in the abundance plots in Figure~\ref{fig:device_abundance_comparsion}, as representatives of gates solving a linearly separable and inseparable problem, respectively. Due to reasons of continuity it is expected that in the 5D space of control voltages there exist well-defined regions where the AND or XOR gate fitness is higher than a threshold \mbox{$F_\mathrm{min}$}. The hypervolume of a specific region hosting a high-fitness gate realization is denoted as $V_0$ and the average hypervolume of these regions as \mbox{$\langle V_0\rangle $}, while the number of different regions is denoted as \mbox{$N_{\rm gates}$}. In the two extreme cases, the different abundances of AND and XOR gates may be due to a very different number of regions \mbox{$N_{\rm gates}$} of similar average hypervolume \mbox{$\langle V_0\rangle $} or due to a similar number of regions with very different average hypervolume.

We determined the reason for the different abundances of AND and XOR gates in the following way. We randomly choose for both devices D1 and D2 an AND gate realization with \mbox{$F > F_{\rm min} = 10$} and an XOR gate realization with \mbox{$F > F_{\rm min} =5$}. In addition, we randomly choose for D1 an XOR gate realization with \mbox{$F > F_{\rm min} =10$} (for D2 we refrained from doing this, because the XOR gate abundance is with this fitness threshold too low for statistical significance).
We assume that the hypervolume $V_0$ of the region hosting the randomly chosen gate realization is representative for all regions hosting gate realizations, so that we do not need to distinguish between $V_0$ and \mbox{$\langle V_0 \rangle$}.
To estimate $V_0$ of each region, we randomly choose control voltages restricted to a local hypercube
with hypervolume $\Delta V$ incorporating the region. Then, we calculate the probability $p_0$ that a combination of $10^4$ randomly chosen control voltages within this local hypercube leads to a gate fitness $F>F_{\rm min}$. This information allows us to estimate $V_0$ as  \mbox{$V_{0}\approx p_0 \Delta V$}. From the global gate abundance $p_{\rm abundance}$, extracted from \mbox{Figure~\ref{fig:device_abundance_comparsion}}, we obtain an estimate of the global hypervolume $V$ of all gate realizations with minimal fitness $F_{\rm min}$ as $V \approx p_{\rm abundance} V_{\rm tot}$, where $V_{\rm tot}=2^5$ V$^5=32$ V$^5$ is the hypervolume of the global hypercube (the control voltage were chosen in a voltage range of 2 V between $-1$ and $1$ V). An estimate of the number of distinct gate realizations with minimal fitness $F_{\rm min}$ in the control voltage space is then found as the ratio between the global hypervolume of all gate realizations and the local hypervolume of a particular gate realization: 
\begin{equation}
  N_{\rm gates}\approx V/V_{0} \approx \frac{p_{\rm abundance} V_{\rm tot} }{p_0 \Delta V} .
\end{equation}
 
\begin{table}
\caption{\label{tab:gate_in_space}
Properties of the realization of AND and XOR gates in the 5D control voltage space for the two devices D1 and D2, for $T=77$ K and $a=5$ nm.}
\begin{ruledtabular}
\begin{tabular}{crllllr}
 & $F_{\rm min}$ & $p_0$ & $p_{\rm abundance}$ & $\Delta V$ [V$^5$] & $V_0$ [V$^5$] & $N_{\rm gates}$ \\
\hline
AND (D1) & $10$  & $0.05$    & $0.0015$  & $0.18$ & 0.0090   & $5$ \\
XOR (D1) & $5$   & $0.0085$  & $0.0005$  & $0.11$  & 0.00094  & $17$ \\
XOR (D1) & $10$  & $0.0012$  & $0.00007$ & $0.11$  & 0.00013  & $17$ \\
AND (D2) & $10$  & $0.0083$  & $0.0015$  & $1.1$   & 0.0091  & $5$ \\
XOR (D2) & $5$   & $0.003$   & $0.0005$  & $0.59$  & 0.0018  & $9$ \\
\end{tabular}
\end{ruledtabular}
\end{table}

The results of these estimates are shown in Table \ref{tab:gate_in_space} for $T=77$ K and $a=5$ nm. For device D1 and \mbox{$F_{\rm min}=10$}, the hypervolume $V_0$ of the XOR gate realization is almost two orders of magnitude lower than that of the AND gate realization ($0.00013$ vs.\ $0.0090$ V$^5$). This shows that for the XOR gate more subtle tuning of the control voltages is required than for the AND gate, an observation that was also made in our previous work when studying the fitness change when changing one of the control voltages \cite{Tertilt2022}. In contrast, the estimated number of distinct realizations $N_{\rm gates}$ of the XOR gate is of comparable magnitude as that of the AND gate (17 vs.\ 5). The smaller abundance of XOR gates as compared to AND gates is thus not due to a smaller number of regions, but due to a much smaller hypervolume of a region. This conclusion is supported by analogous results for device D2. Remarkably, the number $N_{\rm gates}$ of distinct XOR gate realizations for device D1 is not very different  for $F_{\rm min}=5$ and $10$ (by chance even identical: 17). This shows that the strong decrease of the abundance with fitness threshold $F_{\rm min}$ in Figure~\ref{fig:device_abundance_comparsion} is not due to a strong decrease of distinct XOR gate realizations, but due to a strong decrease of the hypervolume of the regions when increasing $F_{\rm min}$, as is also observed when directly comparing the values of $V_0$ ($0.00013$ vs.\ $0.00094$ V$^5$). These observations are crucial for the further development of logic functionality with DNPU technology. 

For the consistency of our choice for the hypervolume $\Delta V$ of the local hypercube in Table \ref{tab:gate_in_space}, two conditions should be fulfilled. (1) The local hypercube should be chosen large enough to contain the region of control voltages with minimal fitness $F_{\rm min}$ hosting the specific gate realization. (2) It should be small enough to avoid overlap with regions hosting other gate realizations. Condition (1) is fulfilled by making sure that at the edges of the local hypercube the gate fitness has decreased well below $F_{\rm min}$, implying $p_0\ll 1$. Condition (2) is fulfilled by making sure that $V_{\rm tot}/\Delta V\gg  N_{\rm gates}$. Both conditions are fulfilled for all the cases in Table \ref{tab:gate_in_space}.

\subsection{\label{sec:main}Temperature dependence\protect}

For practical reasons it is desirable to have DNPU logic functionality  at room temperature instead of 77 K. To investigate potential room-temperature functionality, we have redone all simulations for device D1 at $T = 273$ K. The resulting abundance plots are given in Figure~\ref{fig:abundance_comparsion_highT}. One would expect that nonlinear effects become smaller upon temperature increase, because the relative influence of Coulomb interactions, which are responsible for the nonlinear effects, then becomes less. This is expected to result in a smaller abundance of logical gates, in particular XOR and XNOR gates. This is indeed observed in a comparison of Figure~\ref{fig:abundance_comparsion_highT} with Figure~\ref{fig:device_abundance_comparsion}. Interestingly, the behavior for $T=273$ K and $a=5$ nm is similar as for $T=77$ K and $a=10$ nm. Also, the behavior for $T=273$ K and $a=2.5$ nm is similar as for $T=77$ K and $a=5$ nm. We explain this by a compensation of a decrease of nonlinear effects with increasing temperature and an increase of similar magnitude of nonlinear effects with decreasing hopping distance. We note that for $T=273$ K we could not perform simulations with $a = 1.25$ nm, because the higher temperature allows hops that were almost impossible at $T=77$ K.

 \section{\label{sec:main} Results: Principal component analysis\protect}

 \subsection{\label{sec:main} Analysis of KMC data\protect}

\begin{figure*}[t]
    \includegraphics[width=0.49\textwidth]{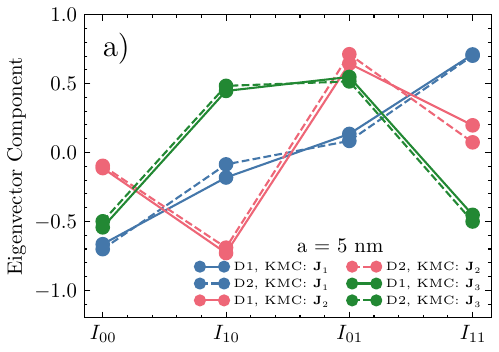}
    \includegraphics[width=0.49\textwidth]{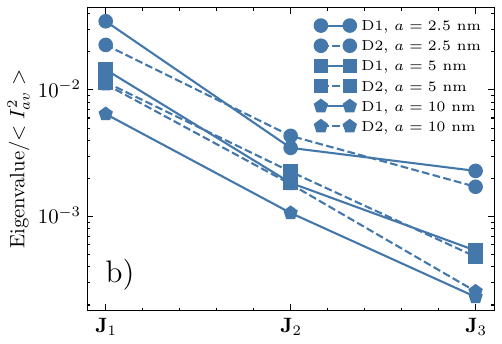}
    \caption{\label{eigenproperties} Left: eigenvectors $\textbf{J}_1$, $\textbf{J}_2$, $\textbf{J}_3$ of the PCA covariance matrix $C$ of the two devices D1 and D2, for $T=77$ K and $a=5$ nm. Right: corresponding normalized eigenvalues, with the results for $a=2.5$ and $10$ nm added.}
\end{figure*}

The results for the eigenvectors $\textbf{J}_i$ $(i=1,2,3)$  of the PCA covariance matrix Eq.~(\ref{eq:covariancematrix}) for $T=77$ K and $a=5$ nm are shown in the left panel of Figure~\ref{eigenproperties} for both devices D1 and D2. Since by construction $\mathbf{J}_0=\mathbf{v}_0=\frac{1}{2}(1,1,1,1)$, we do not show $\mathbf{J}_0$. Remarkably, we find to an excellent approximation $\mathbf{J}_3\approx\mathbf{v}_3=\frac{1}{2}(-1,1,1,-1)$. As discussed above, the equality follows if no correlations between $X$ and $M_{\rm l}$ or $M_{\rm r}$ would be present. The approximate equality suggests that these correlations are indeed small, as we will explicitly verify below. Furthermore, we find $\mathbf{J}_1\approx \mathbf{v}_1=\frac{1}{\sqrt{2}} (0,-1,1,0)$ and $\mathbf{J}_2\approx \mathbf{v}_2=\frac{1}{\sqrt{2}} (-1,0,0,1)$ ($\frac{1}{\sqrt{2}}=0.7071\ldots$). This suggests that there is a high but not perfect l-r symmetry of the devices.

The corresponding eigenvalues $\lambda_i$ are shown in the right panel of Figure~\ref{eigenproperties}, where also results for $a=2.5$ and $10$ nm have been added. For a proper comparison all eigenvalues are divided by $\langle I_{\rm av}^2 \rangle$ as normalization. For all considered cases the normalized eigenvalues are considerably smaller than $1$. This implies that the average current for the different input combinations is by far the dominant quantity, while changes in the current for the different input combinations can be regarded as relatively small modulations. For the same hopping distance $a$ the normalized eigenvalues are similar within a factor of less than two for the two devices, so that the devices display statistically similar behavior. Furthermore, the normalized eigenvalues of both devices show a significant increase with decreasing $a$. In particular, the third normalized eigenvalue strongly increases with decreasing $a$. According to our above analytical solution $\lambda_3 = 4 \sigma^2(X)$ when correlations between $X$ and $M_{\rm l}$ or $M_{\rm r}$ are neglected, this observation suggests a considerable increase of $\sigma^2(X)$ with decreasing $a$. 

\begin{figure*}[t]
    \includegraphics{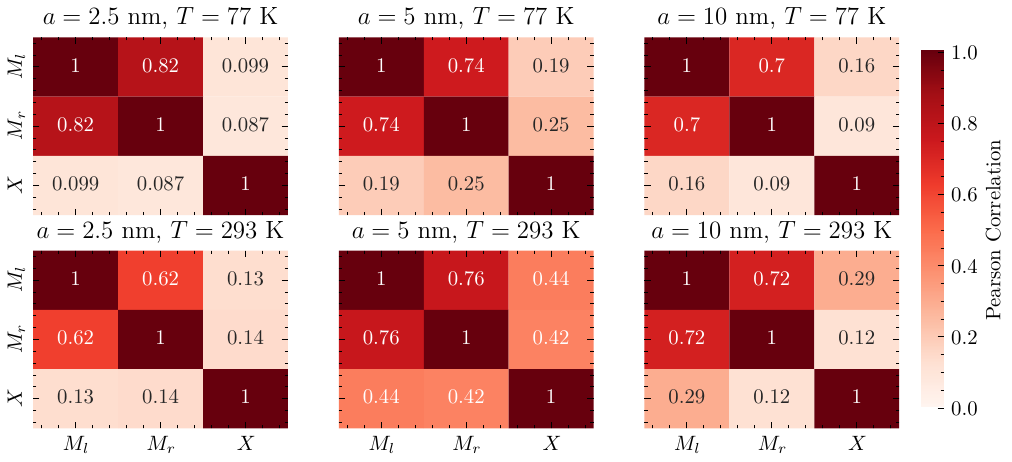}
    \caption{\label{fig:corr_plots}Pearson correlation coefficients for device D1 among $M_{\rm l}$, $M_{\rm r}$ and $X$ for $T=77$ and $293$ K, and $a=2.5$, $5$, and $10$ nm.}
\end{figure*}

In Figure~\ref{fig:corr_plots} we show the Pearson correlation coefficients among the variables $M_{\rm l}$, $M_{\rm r}$ and $X$. The low numbers for the corresponding correlation coefficients confirms the above assumption that the correlation between $X$ and $M_{\rm l}$ or $M_{\rm r}$ is very small, in particular for $T = 77$ K.  Furthermore, we find a considerable correlation between $M_{\rm l}$ and $M_{\rm r}$ with only little dependence on $T$ and $a$. As argued above, a large correlation between $M_{\rm l}$ and $M_{\rm r}$ is important for a realization of high-fitness logic gates, for which $I_{01}\approx I_{10}$. 
As shown in Figure~\ref{fig:abs_parameter_values} of the Supplemental Material \cite{supp} the variances of $M_{\rm l}$ and $M_{\rm r}$ are very similar, which explicitly rationalizes why the eigenvectors $\mathbf{J}_i$ are so close to the $\mathbf{v}_i$.

\subsection{\label{sec:comp_exp}Comparison with experiments\protect}

\begin{figure*}
    \centering
    \includegraphics[width=0.49\textwidth]{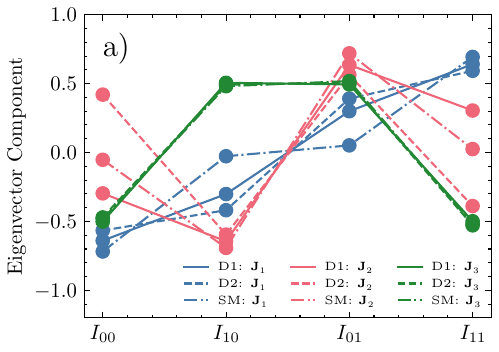}
    \includegraphics[width=0.49\textwidth]{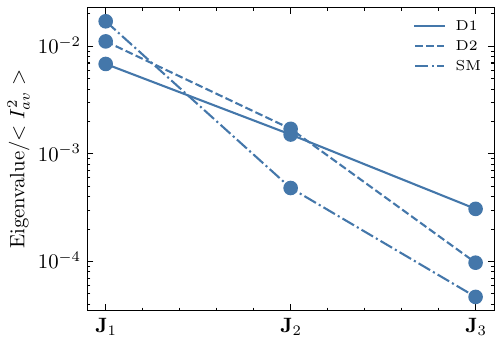}
    \caption{\label{fig:dnn_kmc_eigenproperties}
     Left: eigenvectors $\textbf{J}_1$, $\textbf{J}_2$, $\textbf{J}_3$ of the PCA covariance matrix $C$ of the simulated device D1 and the surrogate model (SM) of a physical device, for $T=77$ K. Right: corresponding normalized eigenvalues, with the results for device D2 added. The KMC results were obtained for $a=5$ nm.}
\end{figure*}

We now compare results of the KMC simulations with experimental results, as emerging from the surrogate model (SM) of a physical device \cite{RuizEuler2020}. The comparison is performed on the level of the properties derived from the PCA. We remind the reader that in order to have similar ranges of voltages we have chosen a smaller variation of input and control voltages in the KMC simulations: input voltages $U_2,U_3\in (0,0.1)$ V, control voltages $U_1,U_4,U_5\in [-0.5,0.5]$ V and control voltages  $U_6,U_7 \in [-0.3,0.3]$ V. As a result, the results are quantitatively different from what is reported above. As before, the results presented here are based on a sampling of $10^4$ randomly chosen control voltage combinations for both simulated devices D1 and D2 as well as the SM of the physical device.   

As seen in the left panel of Figure~\ref{fig:dnn_kmc_eigenproperties}, there is
a fair agreement between the KMC eigenvectors and the SM eigenvectors. We have to an excellent approximation $\mathbf{J}_3\approx\mathbf{v}_3$ both for the simulated devices D1, D2 and the SM of the physical device. For the other two eigenvectors, there is less agreement between $\mathbf{J}_1$ and $\mathbf{v}_1$, $\mathbf{J}_2$ and $\mathbf{v}_2$. As already argued in Sect.~\label{sec:PCA-E} and seen from the comparison of devices D1 and D2 in Figure~\ref{eigenproperties}, the latter two eigenvectors are more susceptible to details of the specific dopant distribution, in particular with respect to the closeness to l-r symmetry. The approximation is therefore less accurate than for $\mathbf{J}_3$. In fact, when calculating the standard deviation ratios of $M_l$ and $M_r$, we get a value of $\frac{\sigma(M_l)}{\sigma(M_r)} = 0.58$ for the first Device and $\frac{\sigma(M_l)}{\sigma(M_r)} = 0.43$ for the second device, indicating l-r asymmetry. In contrast, the surrogate model exhibits closer l-r symmetry, with $\frac{\sigma(M_l)}{\sigma(M_r)} = 0.9$, going along with a closer similarity of  $\mathbf{J}_0$ and $\mathbf{v}_0$.

Comparison of the eigenvalues in the right panel of Figure~\ref{fig:dnn_kmc_eigenproperties} shows that, like for the simulated devices D1 and D2, the largest SM eigenvalue $\lambda_1$ is two orders of magnitude smaller than $\left<I_{\rm av}^2\right>$ and the smallest SM eigenvalue $\lambda_3$ approximately four orders of magnitude smaller.  The agreement shows that the degree of nonlinearity in the simulations and in experiment is very similar, both in terms of the NDR as well as the cross-correlation.
The major difference between the KMC and SM results is the ratio of the first and the second eigenvalue. As seen from the analytical solution Eq.~(\ref{eq:eigenvalues}), the ratio $\lambda_1/\lambda_2=(1 + {\rm Corr}(M_{\rm l},M_{\rm r}))/(1 - {\rm Corr}(M_{\rm l},M_{\rm r}))$ and this ratio is therefore a measure of the correlation between $M_{\rm l}$ and $M_{\rm r}$. We find Pearson correlation coefficients ${\rm Corr}_{\rm D1}(M_{\rm l},M_{\rm r}) = 0.468$, ${\rm Corr}_{\rm D2}(M_{\rm l},M_{\rm r}) = 0.348$, and ${\rm Corr}_{\rm SM}(M_{\rm l},M_{\rm r}) = 0.945$ for D1, D2, and the SM, in agreement with the ratios found in Figure~\ref{fig:dnn_kmc_eigenproperties}. We note that the experimental uncertainty in the number of dopants in the active region in between the electrodes is large and that our modeling of the electrodes as circular segments is very approximate. Considering these and other uncertainties and approximations, the agreement between our simulated results and the experimental results is remarkable.

\section{\label{sec:main}Results: Nonlinearity indicators\protect}

\begin{table*}
    \caption{\label{tab:decomp_values}
    First moment (in nA)  and variance (in nA$^2$) of $M_{\rm l}$, $M_{\rm r}$, and $X$ for $T=77$ and $293$ K, and $a = 5$ nm.}
    \begin{ruledtabular}
    \begin{tabular}{cccccccc}
     Device & Temperature& $\langle M_{\rm l}\rangle$ & $\langle M_{\rm r} \rangle$ & $ \langle X \rangle $ & $\sigma^2(M_{\rm l})$ & $\sigma^2(M_{\rm r})$ & $\sigma^2(X)$ \\
    \hline
    D1 & 77 K   & $0.01496$  & $0.01617$  & $0.00235$  & $0.05218$  & $0.03667$ & $0.01202$\\
    D2 & 77 K   & $0.05431$  & $0.04739$  & $-0.00009$ & $0.04468$  & $0.03800$ & $0.01098$\\
    D1 & 293 K  & $0.04055$  & $0.02641$  & $0.00298$  & $0.04758$  & $0.02764$ & $0.00529$\\
    \end{tabular}
    \end{ruledtabular}
\end{table*}

\begin{figure}
    \includegraphics[width=0.45\textwidth]{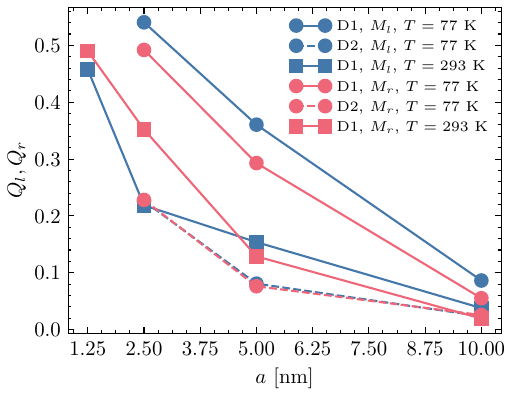}
    \caption{\label{tanh_of_f_g} Nonlinearity indicators $Q_{\rm l}$ and $Q_{\rm r}$ at $T=77$ K (devices D1 and D2) and 293 K (D1) for different $a$.
    }
\end{figure}

We now come to the final analysis level, which is based on the distributions of $M_{\rm l}$, $M_{\rm r}$, and $X$.  The first moments and the variances of these quantities at $T=77$ K (devices D1 and D2) and $T= 293$ K (device D1) for $a = 5$ nm are shown in Table \ref{tab:decomp_values}.
The corresponding left and right NDR indicators $Q_{\rm l}$ and $Q_{\rm r}$, given by Eq.~(\ref{eq:QlQr}), are displayed in Figure~\ref{tanh_of_f_g} for different $a$. Only minor differences are seen between $Q_{\rm l}$ and $Q_{\rm r}$, in agreement with approximate l-r symmetry. There is a considerable difference, occurring by chance, between the two devices, with $Q_{\rm l}$ and $Q_{\rm r}$ for D1 larger than for D2. This difference indicates that the NDR is more pronounced for D1 than for D2.
This is in agreement with the larger abundance of NAND and NOR gates in D1 than in D2 as observed in Figure~\ref{fig:device_abundance_comparsion} for small and intermediate fitness thresholds. Additionally, we observe a very strong increase of $Q_{\rm l}$ and $Q_{\rm r}$ with decreasing hopping distance $a$, indicating that NDR becomes more pronounced when $a$ is small. Indeed, as seen in Figure~\ref{fig:device_abundance_comparsion}, the probability of NAND and NOR gates is strongly enhanced for decreasing $a$. We see from the results for device D1 that increasing the temperature from $T=77$ to 293 K has the same effect as increasing $a$, which agrees exactly with the observation made when comparing the NAND and NOR abundance plots in Figs.~\ref{fig:device_abundance_comparsion} and \ref{fig:abundance_comparsion_highT} for these temperatures.

\begin{figure}
    \includegraphics[width=0.45\textwidth]{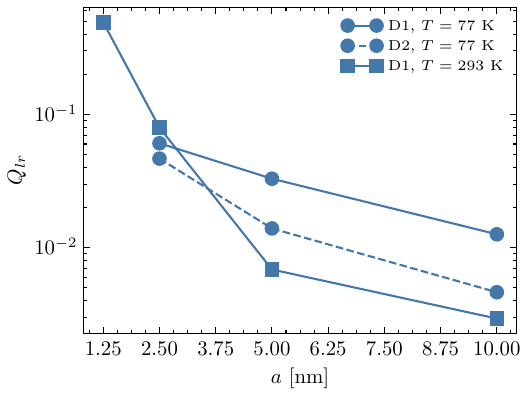}
    \caption{\label{fig:epsilon_dependency} Nonlinearity indicator $Q_{\rm lr}$ at $T=77$ K (devices D1 and D2) and 293 K (D1) for different $a$.
    }
\end{figure}

The  indicator $Q_{\rm lr}$ for the nonlinear coupling between the two inputs, given by Eq.~(\ref{eq:Qlr}), is shown in Figure~\ref{fig:epsilon_dependency}. 
Also this indicator strongly increases upon decreasing the hopping distance $a$. For $a = 2.5$ nm the results for both devices are very similar. Indeed, as seen in Figure~\ref{fig:device_abundance_comparsion}, the abundances of XOR and XNOR gates for $a = 2.5$ nm are also very similar for both devices. For larger $a\ge 5$ nm, $Q_{\rm lr}$  is larger for device D1 than for D2. This is again reflected by the higher occurrence likelihood of XOR and XNOR gates for D1 than D2 in the abundance plots  of Figure~\ref{fig:device_abundance_comparsion}. Also the considerable decrease of $Q_{\rm lr}$ with increasing temperature shows up  when comparing the XOR and XNOR abundance plots in Figs.~\ref{fig:device_abundance_comparsion} and \ref{fig:abundance_comparsion_highT}.

Finally, we mention that we observed from Figure~\ref{fig:corr_plots} that ${\rm Corr}(M_{\rm l},M_{\rm r})$ is large and only weakly dependent on the hopping distance $a$ and temperature $T$. This is compatible with $I_{01}\approx I_{10}$, which is a condition for high-fitness Boolean gates. However, it is not a sufficient condition. The three nonlinearity indicators $Q_{\rm l}$, $Q_{\rm r}$, and $Q_{\rm lr}$ sensitively depend on $a$ and $T$ and are much better measures for the occurrence of high-fitness gates.    

\section{\label{sec:supp} Summary, conclusions and outlook}

We have focused in this work on the critical nonlinear aspects of hopping transport in disordered dopants networks (DNPUs) used in reconfigurable logic. We considered DNPUs with eight electrodes: one output electrode, two symmetrically positioned input electrodes and five control electrodes. From kinetic Monte Carlo (KMC) simulations of the hopping transport, taking into account Coulomb interactions between the charges, the output currents for different voltages applied at the input and control electrodes can be calculated. This allowed us to assess the occurrence of Boolean logic in the five-dimensional (5D) space of control voltages, as quantified by a Boolean gate fitness value of the four-dimensional (4D) current vector for the different input voltages corresponding to the `01', `10', `01', and `00' logic input combinations.

First, we calculated the abundance plots of the six basic Boolean gates from a random hypercube sampling of the 5D control voltage space. For a typical hopping distance of 5 nm the abundance plots for two simulated devices were found to agree well with experimental results \cite{Chen2020} at liquid nitrogen temperature of 77 K. We came to the important conclusion that a small hopping distance or a low temperature is beneficial for the occurrence of high-fitness gates, because nonlinear effects due to the Coulomb interactions between the charges are then stronger than for a large hopping distance or high temperature. 

In a next step, we used a principal component analysis (PCA) to characterize the distribution of the current vectors in more detail. We found that the properties of the eigenvectors of the PCA matrix strongly depend on the degree of symmetry of the dopant network. The corresponding normalized eigenvalues provide a simple representation of the statistical properties of the DNPU. We found a fair agreement between the eigenvectors and the normalized eigenvalues of two simulated devices and a deep neural network (DNN) surrogate model (SM) of a physical device. This shows that our modeling at least qualitatively captures the underlying physics of the DNPUs. It is important to note that, in contrast to other applications of the PCA, all eigenvectors are of key importance. When omitting, e.g., the direction along the eigenvector with the smallest eigenvalue, one would no longer be able to assess the occurrence of XOR and XNOR gates, because of the missing information about the cross-correlation between the two inputs, contained in this eigenvalue.

Finally, we defined three dimensionless nonlinearity indicators $Q_{\rm l}$, $Q_{\rm r}$, and $Q_{\rm lr}$, where $Q_{\rm l}$ and $Q_{\rm r}$ are indicators for negative differential resistance (NDR) with respect to the left and right input, important for the realization of NAND and NOR gates, and $Q_{\rm lr}$ is an indicator for nonlinear coupling between the left and right input, important for the realization of XOR and XNOR gates. On this deepest analysis level, important new insights about the impacts of the hopping length, the temperature, and cross-correlations on the logic functionality were gained.

In addition to the statistical properties obtained from the hypercube sampling, we considered the spatial structure of Boolean gate realizations in the 5D control voltage space. We found the surprising result that for AND and XOR gates, as representatives, respectively, of linearly separable and linearly inseparable gates, the number of regions hosting high-fitness gates is similar, despite the fact that the abundance of AND gates is much higher. This is explained by the much smaller hypervolume of the regions, and the resulting higher sensitivity of the fitness when varying control voltages of  XOR gates than AND gates. 

Different further applications of the presented methodology are conceivable: (1) In this work, we have modified the hopping distance. The physically relevant quantity is the ratio of the typical nearest-neighbor distance of dopants and the hopping distance. The hopping distance is difficult to change without using a different dopant-semiconductor combination, but the distance between the dopants can easily be changed by changing the dopant density. It would therefore be of interest to make a comparison with DNPUs made with a different dopant density. A lower dopant density may increase the relative importance of Coulomb interactions, with beneficial effects for the logic functionality. (2) The proposed decomposition of the current vectors is very straightforward and not dependent on the specific underlying physical realization. Thus, it could be easily adjusted to situations with, e.g., three input electrodes, or where the current vector results from different realizations of reconfigurable logic, such as nanoparticle networks \cite{Bose2015}, or where other device properties such as the size are varied. Work along this line is in progress. (3) A very interesting application of DNPUs is the processing of time-dependent signals. For that case, e.g., the quantification of the cross-correlation may be very helpful to characterize the mixing of signals caused by voltage changes of different input electrodes. (4) Different realizations of the DNPs can display significant device-to-device fluctuations as also observed in other neuro-inspired computing systems \cite{Sebastian2018}. The nonlinearity indicators, introduced above, may yield direct information about the properties and the consequences of these flucations on the device behavior.

\section{\label{sec:acknowledgement} Acknowledgement} This work was funded by the Deutsche Forschungsgemeinschaft (DFG, German Research Foundation) through project 433682494--SFB 1459. We thank \hbox{Dr. Unai Alegre-Ibarra} for setting up the GitHub repository to make the KMC code publicly available (https://github.com/MUTUEL). 

\bibliography{main}



\setcounter{figure}{0}
\renewcommand{\thefigure}{S\arabic{figure}}

\externaldocument{dopant_paper}

\onecolumngrid
\section{Supplemental Material}

\begin{figure*}[h!]
    \includegraphics{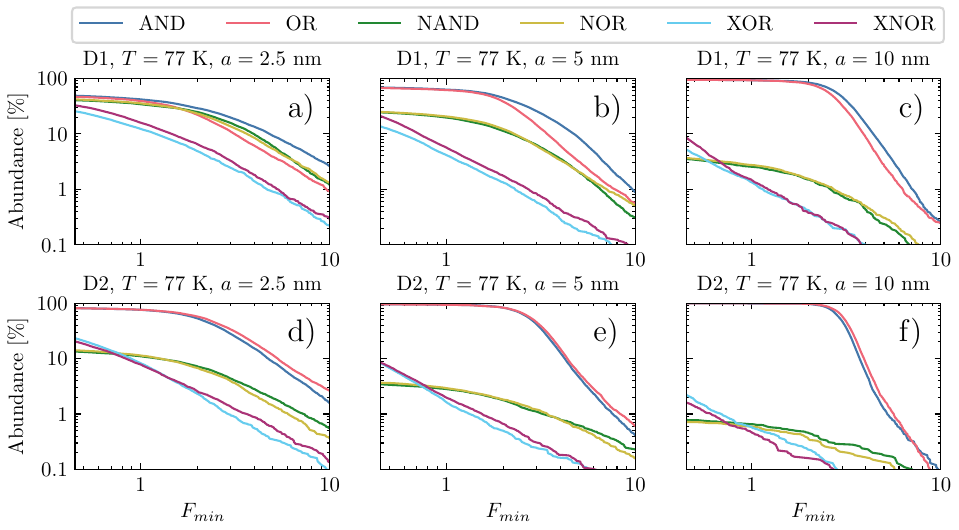}
    \caption{\label{fig:device_abundance_comparsion_k}
    Abundance at $T = 77$ K of Boolean gates with fitness higher than a fitness threshold $F_\mathrm{min}$ in the 5D space of control voltages for the two devices D1 and D2 and hopping distance $a=2.5$, $5$ and $10$ nm. In contrast to Fig.~\ref{fig:device_abundance_comparsion}, for which $k=0.01$, the fitness function Eq.~(\ref{eq:fitnessfunction}) is in this case evaluated with $k=0$.}
\end{figure*}

\begin{figure}[h!]
\includegraphics[width=0.45\textwidth]{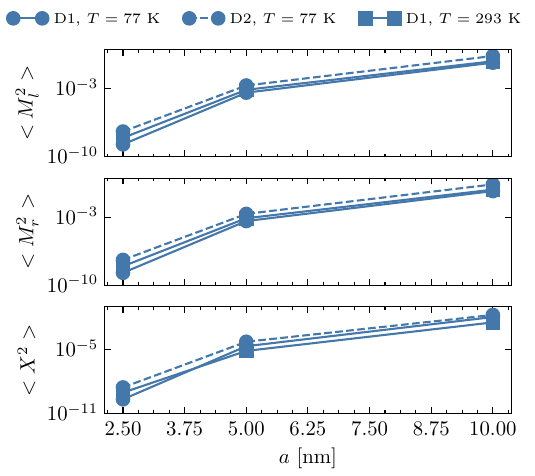}
\caption{\label{fig:abs_parameter_values} Variances $\left<M_{\rm l}^2\right>$, $\left<M_{\rm r}^2\right>$, and $\left<X^2\right>$ for devices D1 and D2 and different hopping distance $a$ at temperatures $T=77$ K (D1 and D2) and $293$ K (only D1).}
\end{figure}


\end{document}